\documentclass[12pt]{article}
\usepackage{graphicx}
\usepackage{url}
\usepackage{hyperref}
\usepackage{epstopdf}
\usepackage[printonlyused]{acronym}
\usepackage{subfig}
\usepackage{amssymb}
\usepackage{amsmath}
\usepackage{amsfonts}

\DeclareMathOperator*{\median}{median}
\usepackage[utf8]{inputenc} 
\usepackage{geometry}
\title{Coactivated Clique Based Multisource Overlapping Brain Subnetwork Extraction}
\author{Chendi Wang, Rafeef Abugharbieh}  
\date{}

\begin{document}



\maketitle              

%
%

\newcommand{\NA}{\textsc{n/a}}	
\newcommand{\eg}{e.g.,\ }	
\newcommand{\ie}{i.e.,\ }	
\newcommand{\etal}{\emph{et al.}}

\newcommand{\file}[1]{\texttt{#1}}
\newcommand{\class}[1]{\texttt{#1}}
\newcommand{\latexpackage}[1]{\href{http://www.ctan.org/macros/latex/contrib/#1}{\texttt{#1}}}
\newcommand{\latexmiscpackage}[1]{\href{http://www.ctan.org/macros/latex/contrib/misc/#1.sty}{\texttt{#1}}}
\newcommand{\env}[1]{\texttt{#1}}
\newcommand{\BibTeX}{Bib\TeX}

\DeclareUrlCommand\DOI{}
\newcommand{\doi}[1]{\href{http://dx.doi.org/#1}{\DOI{doi:#1}}}

\newcommand{\webref}[2]{\href{#1}{#2}\footnote{\url{#1}}}

\makeatletter
\newenvironment{epigraph}{%
	\begin{flushright}
	\begin{minipage}{\columnwidth-0.75in}
	\begin{flushright}
	\@ifundefined{singlespacing}{}{\singlespacing}%
    }{
	\end{flushright}
	\end{minipage}
	\end{flushright}}
\makeatother

\newcommand{\FIXME}[1]{\typeout{**FIXME** #1}\textbf{[FIXME: #1]}}


\setcounter{page}{1}
\pagenumbering{arabic}
\begin{center}
Biomedical Signal and Image Computing Lab, UBC, Canada

chendiw@ece.ubc.ca, rafeef@ece.ubc.ca
\end{center}
\begin{abstract}
Subnetwork extraction using community detection methods is commonly used to study the brain's modular structure. Recent studies indicated that certain brain regions are known to interact with multiple subnetworks. However, most existing methods are mainly for non-overlapping subnetwork extraction.
In this paper, we present an approach for overlapping brain subnetwork extraction using cliques, which we defined as co-activated node groups performing multiple tasks. We proposed a multisource subnetwork extraction approach based on the co-activated \textit{clique}, which (1) uses task co-activation and task connectivity strength information for clique identification, (2) automatically detects cliques of different sizes having more neuroscientific justifications, and (3) shares the subnetwork membership, derived from a fusion of rest and task data, among the nodes within a clique for overlapping subnetwork extraction. 
On real data, compared to the commonly used overlapping community detection techniques, we showed that our approach improved subnetwork extraction in terms of group-level and subject-wise reproducibility. 
We also showed that our multisource approach identified subnetwork overlaps within brain regions that matched well with hubs defined using functional and anatomical information, which enables us to study the interactions between the subnetworks and how hubs play their role in information flow across different subnetworks. We further demonstrated that the assignments of interacting/individual nodes using our approach correspond with the posterior probability derived independently from our multimodal random walker based approach.

Keywords: Clique, Overlapping Brain Subnetwork Extraction, Multisource Fusion, Functional Connectivity, Hypergraph
\end{abstract}
\section{Introduction}
The mainstream of brain subnetwork extraction and standard definition of modularity focus on nonoverlapping definition. However, studies have shown evidences of the existence of overlapping brain subnetworks, hence the methods for nonoverlapping subnetwork extraction are limited by neglecting inclusive relationships \cite{Ferrarini-2009-Hier}. There are emerging approaches for discovering overlapping modular network structure, which implies that single nodes may belong in more than one specific module. We here summarize some representative approaches used in brain subnetwork extraction application, and detailed information can be found in a review paper on general overlapping community detection \cite{Xie-2013-Overlapping} . 

The \ac{CPM} is one of the earliest methods for overlapping community detection \cite{Palla-2005-Uncovering}. It is based on the assumption that communities tend to be comprised of overlapping sets of cliques, \ie fully connected subgraphs. It identifies overlapping communities by searching connected cliques. First, all cliques of a fixed size $k$ must be detected, and a clique adjacency matrix is constructed by taking each clique as a vertex in a new graph. Two cliques are considered connected if they share $k$-1 nodes. Communities are detected corresponding to the connected components of the clique adjacency matrix. Since a vertex can be in multiple cliques simultaneously, mapping the communities from the clique level back to the node level may result in nodes being assigned to multiple communities \cite{Xie-2013-Overlapping, Sporns-2016-Modular}.
The limitation of \ac{CPM} is that it operates on binarized graph edges, thus cannot handle weighted graphs \cite{Yoldemir-2016-Multimodal}.

A new definition of modularity has been proposed to discover the overlapping subnetwork based on unbiased cluster coefficients using resting state connectivity \cite{Ferrarini-2009-Hier}. However, methods based on the modularity function $Q$ suffer from degenerate partitions and resolution limit \cite{Sporns-2016-Modular}. 
Another line of studies is to transform a network into its corresponding line graph, where the nodes represent the connections in the original network. Thus, the nonoverlapping community detection (modularity maximization used in \cite{Evans-2009-Line} and agglomerative hierarchical clustering used in \cite{Ahn-2009-Link}) on the line graph will result in overlapping subnetworks in the original network. There exist inherent limitations in the nonoverlapping community detection used for the line graph (resolution limit for modularity maximization and local sub-optimum for hierarchical clustering).

Fuzzy community detection algorithms quantify the strength of association between all pairs of communities and nodes \cite{Xie-2013-Overlapping}. Fuzzy k-menas clustering \cite{Zhang-2007-Identification} and fuzzy affinity propagation \cite{Ding-2010-Overlapping}
have been applied to detect overlapping brain subnetwork extraction. However, one has to use an \textit{ad hoc} threshold for extracting interacting nodes or independent nodes from the membership vector.

Local expansion and optimization algorithms grow a \textit{natural} community \cite{Lancichinetti-2009-Detecting} or a partial community based on local benefit functions \cite{Xie-2013-Overlapping}. 
One example is \ac{CIS}, which has been explored for brain subnetwork extraction \cite{Yan-2011-Detecting}. Taking each node as a partial subnetwork, \ac{CIS} expands the subnetwork by determining if any other nodes belong to this existing subnetwork using a local function to form a densely connected group of nodes. Its limitation is the sensitivity to a density factor that controls subnetwork size \cite{Yoldemir-2016-Stable}. Another good example is the \ac{OSLOM} \cite{Lancichinetti-2011-Finding}, which uses statistical significance of a subnetwork when tested against a global randomly generated null model during community expansion. \ac{OSLOM} has been shown to outperform many state-of-the-art community detection techniques.

In a previous work from our lab, the \ac{RD} concept from theoretical biology for modeling the evolution of interacting and self-replicating entities was used to identify subnetworks. Further, the \ac{RD} formulation was extended to enable overlaps between subnetworks by incorporating a graph augmentation strategy \cite{Torsello-2008-Beyond}. This approach, \ac{SORD} \cite{Yoldemir-2016-Stable}, has demonstrated its superiority over many commonly used overlapping subnetwork extraction methods, including \ac{OSLOM}.

Most of the algorithms aforementioned are based on one single source, such as resting state functional connectivity. \ac{CSORD}, the multimodal version of \ac{SORD}, is one of the few overlapping methods which considers multi-source information. \ac{CSORD} is based on survival probabilities of different genders in evolution and graph augmentation \cite{Torsello-2008-Beyond}. However, its theoretical background for overlapping assumption based on graph augmentation has relatively indirect neuroscientific justifications. We here explore the direction of integrating multisource information for the overlapping subnetwork extraction by using the straightforward \textit{clique} concept. 

\section{Co-activated Clique Based Multisource Overlapping Subnetwork Extraction}
\begin{figure}
	\centering
    \includegraphics[width=5in]{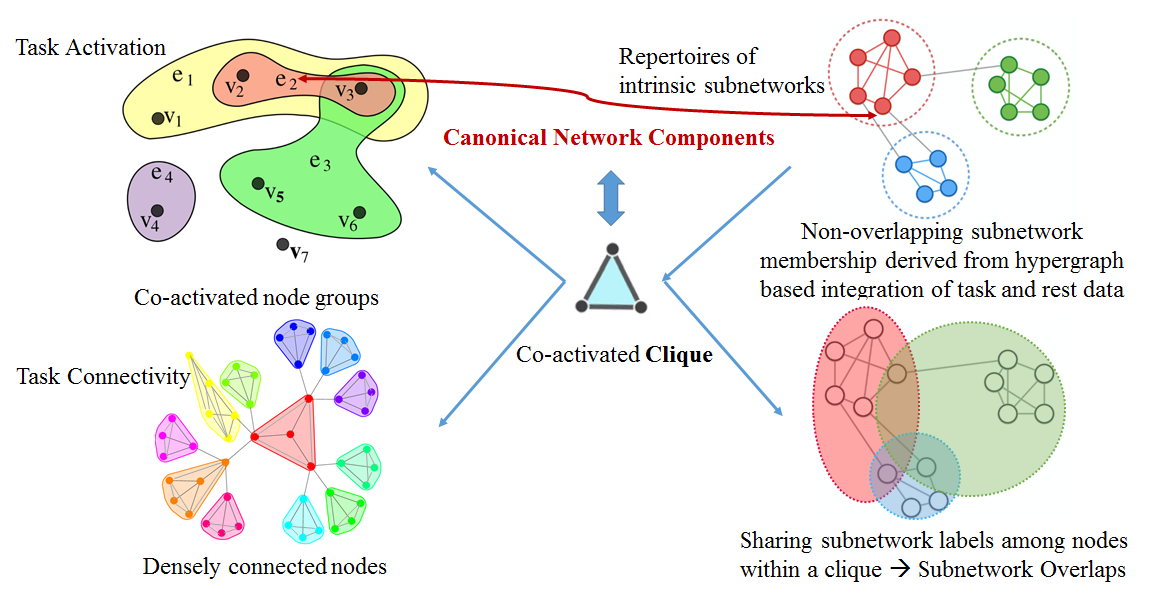}
    \caption{The schematic illustration of multisource clique based overlapping subnetwork extraction approach.}
    \label{fig:MCSE_workflow}
\end{figure}

Recent study has indicated that repeatedly activated nodes in different tasks could be canonical network components in the pre-existing repertoires
of intrinsic subnetworks \cite{Park-2014-Graph}, we argue that the \textit{clique} concept closely resembles groups of nodes which are the \textit{canonical network components}. Based on the basic observation that typical communities consist of several cliques that tend to share many of their nodes \cite{Palla-2005-Uncovering}, clique-based approach would be a straightforward way to find overlapping brain subnetworks. However, the existing clique-based subnetwork extraction approach \ac{CPM} (kclique) \cite{Palla-2005-Uncovering} 
has three major limitations that it can only handle binary graphs, but not weighted graphs;  the size $k$ of cliques is fixed, which needs to be adjusted for different types of networks; and it only uses uni-source information. In order to tackle the aforementioned limitations, we here propose a multisource subnetwork extraction approach based on co-activated \textit{clique}, which (1) uses task co-activation and task connectivity strength information for clique identification, (2) automatically detects cliques with different sizes having more neuroscientific justifications, and (3) shares the subnetwork membership, derived from multisource hypergraph based approach we recently proposed \cite{Wang-2017-Hyper}, among nodes within a clique for overlapping subnetwork extraction. The schematic illustration of our approach is shown in \autoref{fig:MCSE_workflow}.

We first detect co-activated groups of brain nodes across different tasks based on an activation fingerprint idea, and then identify densely connected cliques based on task-induced weighted connectivity. Core cliques are further detected using clique properties we defined.
The nodes within a clique should belong to the same subnetworks due to the close relationship between nodes in a fully connected clique, we thus share the subnetwork membership of nodes within a clique to facilitate overlapping subnetwork assignment. The initial subnetwork membership for each node is derived from non-overlapping subnetwork extraction technique, which is based on the fusion of resting state connectivity and task information embedded with high order relations using hypergraph (see details and notations in \cite{Wang-2017-Hyper}). 
The difference of our approach from the traditional uni-source kclique method is the utilization of both the task co-activation information
and the connectivity weights (only the binarized connectivity is used in kclique method). The co-activated cliques derived using our approach have flexible clique sizes, which has more neuroscientific justifications than the fixed $k$. Besides, we explore if our proposed clique node subnetwork membership sharing idea can generate more straightforward and biologically meaningful results than the existing multisource method \ac{CSORD}.

\subsection{Clique Identification Based on Task Co-activation}
We define cliques as co-activated groups of brain nodes that are densely connected in our approach. We first identify the co-activated groups of brain nodes (coarse cliques) using an activation fingerprint idea. Then we refine coarse cliques into cliques, within which nodes are densely connected to each other based on task-induced connectivity information. We denote the clique set as $CS$, and the coarse clique set as $CCS$.
Given $T$ different tasks, one can construct a hypergraph with an $N\times T$ incidence matrix $\mathbf{H}$, where $N$ is the number of brain regions and $T$ is the number of tasks (hyperedge $e$). $h(v,e) = 1$ when the brain region node $v$ is activated in the task corresponding to hyperedge $e$. 
The task-induced connectivity matrix $\mathbf{C}^{\text{task}}$ is generated by removing all inter-block rest periods from all regions’ time courses and computing pairwise Pearson's correlations of time courses which were concatenated through block/event durations across all the tasks. 
The underlying assumption for our clique identification is that nodes in the same clique should be co-activated across tasks at times from $t = 1\ldots T$, where $t$ indicates the number of tasks, in which the nodes are co-activated. 
There are two steps involved in our clique identification, which (1) pre-selects sets of coarse cliques in all $T$ layers, (2) and refines the coarse cliques into cliques.

The approach starts with a pre-selection of coarse cliques $CCS$, which might include loosely connected nodes that are co-activated. Take each row from the incidence matrix $\mathbf{H}$ as a activation fingerprint vector $f$ corresponding to the task activation pattern of a node. For example, if one node is activated in the 1st, 3rd and 6th out of the seven tasks, the corrsponding $f = [1 0 1 0 0 1 0]$. We next operate \textit{bit-wise and} between the fingerprints from a node pair $\{i, j\}$, which gives us an output fingerprint vector of co-activation patterns $\mathbf{SF}_{ij}$:
\begin{equation}
\label{eq:clique_FP}
\mathbf{SF}_{ij} = f_i \ \land \ f_j, 
\end{equation}
where $f_i$ and $f_j$ are the activation fingerprint vectors of node pair $i$ and $j$, and $\mathbf{SF}$ is the matrix containing the co-activation fingerprint vectors between the nodes in each node pair. We then define a matrix $\mathbf{NT}$ which counts the number of co-activated tasks between two nodes:
\begin{equation}
\label{eq:clique_NT}
\mathbf{NT}_{ij} = \sum_{t=1}^T \mathbf{SF}_{ij}(t), 
\end{equation}
where $\mathbf{SF}_{ij}$ is the co-activation fingerprint vector of length $T$. Next, we define the node set $PS^{=t}$ which contains nodes that are co-activated together for $t$ times as:
\begin{equation}
\label{eq:clique_PS1}
PS^{=t} = \bigcup_{\forall i,j \ s.t. \ \mathbf{NT}_{ij}=t} \{i, j\}, 
\end{equation}
and define the node set $PS^{>t}$ which contains nodes that are co-activated together for greater than 
$t$ times as:
\begin{equation}
\label{eq:clique_PS2}
PS^{>t} = \bigcup_{\forall i,j \ s.t. \ \mathbf{NT}_{ij}>t} \{i, j\}. 
\end{equation}

Based on the definition above, we follow the four steps as below to identify the coarse cliques.

\textbf{Step 1} We extract $M_t$ pre-selected sets of co-activated coarse cliques from the nodes in $PS^{=t}$. We identify $\{CCS_1^{=t}, CCS_2^{=t}, \ldots, CCS_{M_t}^{=t}\}$ by ensuring all the node pairs within a certain set share the same co-activation fingerprint vector in $\mathbf{SF}$:
\begin{equation}
\label{eq:clique_PS1_set_CCS}
CCS_m^{=t} = \{p_1^m, p_2^m,\ldots,p_{N_m}^m \ | \ \exists\ p_i^m, p_j^m \in PS^{=t}, \ s.t.\ \mathbf{SF}_{p_i^m p_j^m} = \mathbf{SF}_{p_1^m p_2^m}\}.
\end{equation}
where $m=1,\ldots,M_t$.
The minimal rank of $CCS_m^{=t}$ is 2, being only one node pair within a coarse clique. The nodes identified in a coarse clique are fully connected to each other defined by sharing the same co-activation pattern. 

\textbf{Step 2} Similarly, we extract $M_t$ extended sets of co-activation coarse cliques, 
$\{CCS_1^{>t}, CCS_2^{>t}, \ldots, CCS_{M_t}^{>t}\}$, from the nodes in $PS^{>t}$, based on the co-activation patterns between nodes in $CCS_m^{=t}$ and $PS^{>t}$: 
\begin{equation}
\label{eq:clique_PS2_set_CCS}
CCS_m^{>t} = \bigcup_{\forall \ i\in CCS_m^{=t}, \ \exists\ j \in PS^{>t},\ s.t.\ \mathbf{SF}_{ij} \land \mathbf{SF}_{p_1^m p_2^m} = \mathbf{SF}_{p_1^m p_2^m}} \{j\}.
\end{equation}
We do not consider the coarse clique set selection for the nodes which only exist in the node set $PS^{>t}$ for the $t$th layer, since those will be selected in the pre-selected sets in $t+1$ layer.

\textbf{Step 3} We then generate $M_t$ coarse clique sets by merging the pre-selected and extended sets together as:
\begin{equation}
\label{eq:clique_CCSt}
\begin{split}
CCS^{t} = \{CCS_1^{t}, CCS_2^{t},\ldots,CCS_{M_t}^{t}\}\\
CCS_m^t = CCS_m^{=t} \cup CCS_m^{>t},\ m = 1,\ldots,M_t
\end{split}
\end{equation}

\textbf{Step 4} Extract the coarse clique set $CCS$ across layers in the order from $T$ to 1:
\begin{equation}
\label{eq:clique_CCS}
CCS = \bigcup_{t=T,\ldots,1} CCS^t.
\end{equation}

The second part of clique identification is to refine the coarse cliques into cliques. When we extract $CCS = \{CCS^{1}, CCS^{2},\ldots,CCS^{M}\}$, there still exist loosely connected nodes in the coarse cliques, mostly from lower layers when $t$ is small, especially when $t=1$. Hence, we subsequently extract cliques based on the strength information from task-induced connectivity matrix $\mathbf{C}^{\text{task}}$ and hypergraph properties. 
We formulate a coarse clique set, $CCS^k = \{p_1^k, p_2^k,\ldots,p_{M_k}^k\}$ where there are $M_k$ nodes within, as an $M_k\times M_k$ simple graph with the weights between nodes being the task-induced connectivity pairwise edge strength. 
We next apply a local thresholding \cite{Wang-2016-Modularity} on the $M_k\times M_k$ connectivity matrix $\mathbf{C}^{\text{task}\mbox{-} k}$ to find out the most closely connected nodes to each node, and binarize the thresholded matrix to generate an adjacency matrix $\mathbf{A}^{\text{task}\mbox{-} k}$. We then transform $\mathbf{A}^{\text{task}\mbox{-} k}$ into its hypergraph $\mathbf{H}^{\text{task}\mbox{-} k}$ using $\mathbf{A} = \mathbf{HWH}^{T} - \mathbf{D_v}$, where the locations with 1 in each hyperedge correspond to the nodes that comprise a fully connected subgraph, \ie cliques $CS^c$. We extract $N_c$ cliques:
\begin{equation}
\label{eq:clique_CS}
CS = \{CS^1, CS^2,\ldots,CS^{N_c}\}.
\end{equation}


\subsection{Clique Property Computation}
We present three properties that can be derived to study the cliques for further network analysis.

(1) Co-activation times $NT^c$ of a clique $CS^c$, \ie the number of ones in the clique co-activation fingerprint:
\begin{equation}
\label{eq:clique_CSF}
CSF^c = \bigwedge_{\forall \ i\in CS^c} f_i, 
\end{equation}
then the co-activation times:
\begin{equation}
\label{eq:clique_NCOA}
NCOA^c = \sum_{t=1}^T CSF^c(t).
\end{equation}


(2) Activation times in a clique:
\begin{equation}
\label{eq:clique_aveNT}
NA^c = \frac{1}{|CS^c|} \sum_{\forall \ p\in CS^c} \sum_{t=1,\ldots, T} f_p(t).
\end{equation}

(3) Clique overlap ratio - the times of a clique overlaps with other cliques divided by the size of a clique, \ie the number of nodes within a clique.
We first define the set of cliques which node $i$ belongs to as a label set: 
\begin{equation}
\label{eq:node_LC}
LC_i = \{c_1^i, c_2^i,\ldots,c_{N_i}^i\}, c_k^i \in 1,\ldots,N_c,
\end{equation}
where $LC_i$ is an empty set when node $i$ does not belong to any cliques. We then define the clique overlap ratio as:
\begin{equation}
\label{eq:clique_NE}
RCO^c = \frac{1}{|CS^c|}|\bigcup_{\forall \ p\in CS^c}LC_p|.
\end{equation}
\subsection{Core Clique Identification}
Based on the clique properties, we further identify core cliques out of clique sets for the future overlapping subnetwork extraction. We argue that core cliques should have relatively high co-activation times, high activation times, and high clique overlap ratio. We then devise a core clique selection criterion based on the combination of the clique properties. We normalize all the property values into the range of [0, 1] by dividing individual values by the maximum across all the cliques. The criterion is set as below:
\begin{equation}
\label{eq:Cclique_criteria}
\rho = \frac{\median_{\forall \ i\in CS}\{NCOA^i\}}{\max_{\forall \ i\in CS}\{NCOA^i\}} + \frac{\median_{\forall \ i\in CS}\{NA^i\}}{\max_{\forall \ i\in CS}\{NA^i\}} + \frac{\median_{\forall \ i\in CS}\{ROC^i\}}{\max_{\forall \ i\in CS}\{ROC^i\}}.
\end{equation}

For any clique $c$ which satisfies the criterion:
\begin{equation}
\label{eq:Cclique_criteria_clique}
\frac{NCOA^c}{\max_{\forall \ i\in CS}\{NCOA^i\}} + \frac{NA^c}{\max_{\forall \ i\in CS}\{NCOA^i\}} + \frac{ROC^c}{\max_{\forall \ i\in CS}\{ROC^i\}} > \rho, 
\end{equation}
it is selected into the core clique set.
%
\subsection{Clique Based Overlapping Subnetwork Extraction}
Based on the identified core cliques, we further deploy a subnetwork membership sharing technique to identify overlapping subnetworks.
The underlying rationale is that the nodes residing within the same clique behave very similarly to perform some basic functions in tasks, thus, they should be within the same subnetworks. 

In a brain graph with $N$ nodes, let $\mathbf{C}^{\text{rest}}$ be an $N\times N$ resting state connectivity matrix, and we have already labeled the non-overlapping subnetwork membership for each node using $\mathbf{C}^{\text{rest}}$. We have also defined the clique membership of a node $i$ as $LC_i$ in \autoref{eq:node_LC}. We then share the subnetwork membership of the nodes within a clique to facilitate overlapping subnetwork assignment.

First, $M_s$ subnetworks are extracted using non-overlapping community detection approach applied on $\mathbf{C}^{\text{rest}}$. We define the subnetwork membership of a node $i$ as:
\begin{equation}
\label{eq:node_label_s}
label(i) = {s},\ i\in 1,\ldots,N,\ s\in 1,\ldots,M_s.
\end{equation}
Next, we deploy a sharing scheme of the subnetwork membership label from $label(i) $ of a node $i$, with the label set of the remaining nodes in the clique where node $i$ belongs to:
\begin{equation}
\label{eq:expand_LS}
LS(i) = \bigcup_{\forall\ c\in LC_i} \bigcup_{\forall\ p\in CS^c} label(p),
\end{equation}
and 
\begin{equation}
\label{eq:expand_label}
label(i) = label(i) \cup LS(i).
\end{equation}

We have also explored replacing the resting state connectivity matrix $\mathbf{C}^{\text{rest}}$ with the multisource connectivity matrix $\mathbf{C}^{\text{t-r}}$ defined in \cite{Wang-2017-Hyper}. We argue that we should further integrate the activation information from task data with high order relation information presented by hypergraph and the rest data when identifying the non-overlapping subnetwork membership.
\section{Experiments}
We first compare our multisource clique based approach against the uni-source kclique method \cite{Palla-2005-Uncovering}, which is the closest straightforward way to identify overlapping subnetworks. Next we compare against \ac{SORD}, which has been proven to outperform the state-of-the-art techniques such as \ac{OSLOM}, and \ac{CSORD} (the multisource version of \ac{SORD}) \cite{Yoldemir-2015-Coupled} to see if our proposed approach have more direct biological intuition for the overlapping subnetwork extraction. We also examine the nodes within subnetwork overlaps derived by our approach by assessing the probability of a node belonging to subnetworks using our recently proposed multimodal \ac{RW} approach \cite{Wang-2017-RW}, to verify that our overlapping subnetwork assignments correspond with the posterior probability.
\subsection{Materials}
We used the resting state \ac{fMRI} and task \ac{fMRI} scans of 77 unrelated healthy subjects from the \ac{HCP} dataset \cite{van-2013-HCP}. Two sessions of resting state \ac{fMRI} with 30 minutes for each session, and 7 sessions of task \ac{fMRI} data were available for multisource integration. The seven tasks are working memory (total time: 10:02), gambling (6:24), motor (7:08), language (7:54), social cognition (6:54), relational processing (5:52) and emotion processing (4:32). 
Preprocessing already applied to the \ac{HCP} \ac{fMRI} data includes gradient distortion correction, motion correction, spatial normalization to \ac{MNI} space with nonlinear registration based on a single spline interpolation, and intensity normalization \cite{Glasser-2013-Minimal}.
Additionally, we regressed out motion artifacts, mean white matter and cerebrospinal fluid confounds, and principal components of high variance voxels using compCor \cite{Behzadi-2007-Component}. Next, we applied a bandpass filter with cutoff frequencies of 0.01 and 0.1 Hz for resting state \ac{fMRI} data. For task \ac{fMRI} data, we performed similar temporal processing, except a high-pass filter at 1/128 Hz was used. The data were further demeaned and normalized by the standard deviation.
We then used the \ac{HO} atlas \cite{Desikan-2006-Automated}, which has 112 \ac{ROI}s, to define the brain region nodes. We chose the well-established \ac{HO} atlas because it sampled from every major brain system, and consists of the highest number of subjects with both manual and automatic labelling technique compared to other commonly used anatomical atlases. Voxel time courses within \ac{ROI}s were averaged to generate region time courses. The region time courses were demeaned, normalized by the standard deviation. Group level time courses were generated by concatenating the time courses across subjects. The Pearson's correlation values between the region time courses were taken as estimates of \ac{FC} matrices. Negative elements in all connectivity matrices were set to zero due to the currently unclear interpretation of negative connectivity \cite{Skudlarski-2008-Measuring}.
For task activation, we applied the activation detection on the seven tasks available following the steps described in \cite{Wang-2017-Hyper}. 

We further applied local thresholding \cite{Wang-2016-Modularity} on $\bar{\mathbf{C}}^{\text{task}}$ by setting graph density to be 0.1 to generate the hypergraph when we identified cliques from the coarse clique set. We selected a relatively strict threshold to only select those most closely connected nodes to form cliques. 0.1 has been chosen based on the cross-validation on inter-subject reproducibility within the range between 0.03 to 0.2 at the interval of 0.01.
The non-overlapping subnetworks were derived from the resting state connectivity matrix $\mathbf{C}^{\text{rest}}$ or multisource connectivity matrix $\mathbf{C}^{\text{t-r}}$ (generated using strategies from \cite{Wang-2017-Hyper}) using \ac{Ncuts}, when the number of subnetworks was set to 7, same as the abaialble number of tasks.
\section{Results}
We compared the overlapping subnetwork extraction using our proposed \ac{MCSE} approach with $\mathbf{C}^{\text{t-r}}$, or $\mathbf{C}^{\text{rest}}$ against the uni-source kclique approach \cite{Palla-2005-Uncovering}, \ac{SORD} \cite{Yoldemir-2016-Stable}, which has been demonstrated to outperform state-of-the-art overlapping community detection methods including \ac{OSLOM} when applied to brain subnetwork extraction, and \ac{CSORD} \cite{Yoldemir-2015-Coupled}, the multisource extension of \ac{SORD}. Two uni-source approaches extract overlapping subnetworks using resting state data. The parameters for kclique were set using the cross-validation on the clique size $k$ based on inter-subject reproducibility from the suggested range $[3,\ldots,6]$ \cite{Palla-2005-Uncovering} and reasonable graph densities from 0.03 to 0.2 at the interval of 0.01. \ac{SORD} and \ac{CSORD} applied 100 bootstraps by sampling with replacement as suggested in \cite{Yoldemir-2016-Stable}.
We also evaluated the probability of a node being assigned to a subnetwork using our recently proposed \ac{RW} based approach \cite{Wang-2017-RW} to examine the proposed clique-based overlapping subnetwork identification.
All statistical comparisons are based on the Wilcoxon signed rank test with significance declared at an $\alpha$ of 0.05 with Bonferroni correction.
\subsection{Comparison with Existing Overlapping Subnetwork Extraction Methods}
We quantitatively evaluated the contrasted approaches based on test-retest reliability and inter-subject reproducibility, since ground truth subnetworks are unknown for the real data of human brain.
\subsubsection{Group-level Subnetwork Extraction Reproducibility}
\begin{figure}
	\centering
    \includegraphics[width=3.5in]{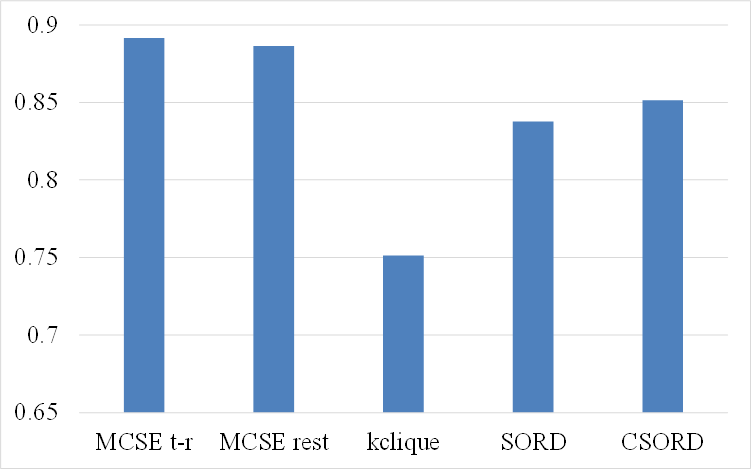}
    \caption{Group-level Subnetwork Extraction reproducibility based on data from two different sessions. \ac{MCSE} outperforms all other contrasted approaches.}
    \label{fig:over_DICE_group}
\end{figure}

We first assessed the test-retest reliability based on group level subnetworks extracted separately from two sessions of rest and task data (each of the seven tasks includes two sessions of \ac{fMRI} data) using \ac{DSC}. 
The subnetworks extracted from the first session's data are taken as the “ground truth”, against which the subnetworks from the second session are compared. 
We found that our proposed \ac{MCSE} outperforms all other contrasted approaches, by achieving a \ac{DSC} between subnetworks extracted from two sessions of data at 0.8917 with $\mathbf{C}^{\text{t-r}}$ and 0.8865 with $\mathbf{C}^{\text{rest}}$, against kclique at 0.7514, \ac{SORD} at 0.8378, and \ac{CSORD} at 0.8514, see \autoref{fig:over_DICE_group}. 
%
\subsubsection{Subject-wise Subnetwork Extraction Reproducibility}
\begin{figure}
	\centering
	    \includegraphics[width=4.8in]{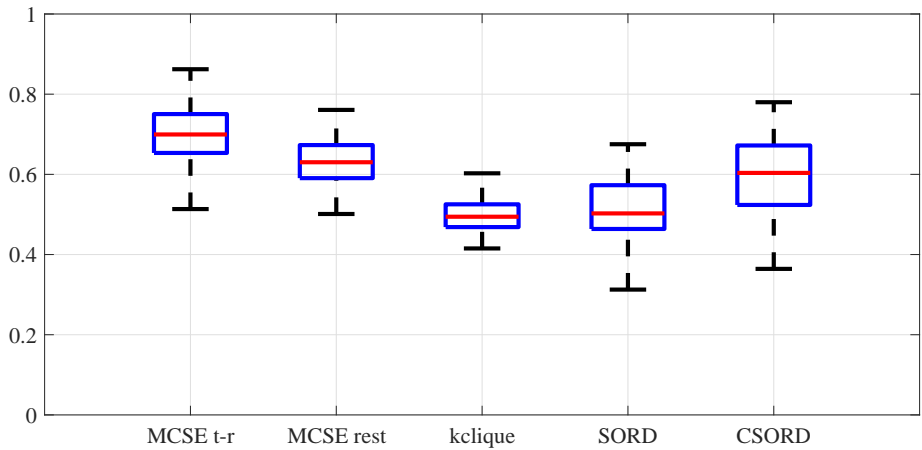}
    \caption{Subject-wise level inter-subject reproducibility of subnetwork extraction. Our proposed \ac{MCSE} approach outperforms existing state-of-the-art overlapping community detection methods.}
    \label{fig:over_repro_subj}
\end{figure}

We assessed the inter-subject reproducibility by comparing the subnetwork extraction results using subject-wise data against the group level data, \autoref{fig:over_repro_subj}. 
The average \ac{DSC} between subject-wise and group level subnetworks across 77 subjects based on five approaches are \ac{MCSE} with $\mathbf{C}^{\text{t-r}}$ at 0.7024$\pm$0.0722, \ac{MCSE} with $\mathbf{C}^{\text{rest}}$ at 0.6281$\pm$0.0583, kclique at 0.4967$\pm$0.0430, \ac{SORD} at 0.5129$\pm$0.0774, and \ac{CSORD} at 0.5952$\pm$0.0901, respectively. \ac{MCSE} with both $\mathbf{C}^{\text{t-r}}$ and $\mathbf{C}^{\text{rest}}$ are found to achieve statistically higher inter-subject reproducibility than constrasted approaches based on the Wilcoxon signed rank test at p$<10^{-10}$ and p$<0.005$. respectively.
Further, \ac{MCSE} with $\mathbf{C}^{\text{t-r}}$ outperforms $\mathbf{C}^{\text{rest}}$ at p$<$0.00001, which confirms the benefit of incorporating the task information embedded with higher order relations in assigning non-overlapping subnetwork membership.

\subsection{Biological Meaning - Analyzing Function Integration}
\begin{figure}
	\centering
	\subfloat[Task activation]{
	    \includegraphics[width=2.4in]{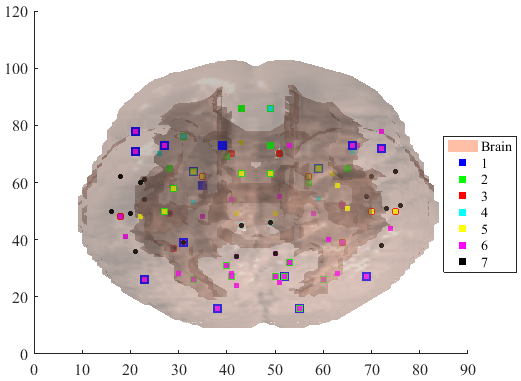}
	}
	\subfloat[\ac{MCSE} with $\mathbf{C}^{\text{t-r}}$]{
	    \includegraphics[width=2.4in]{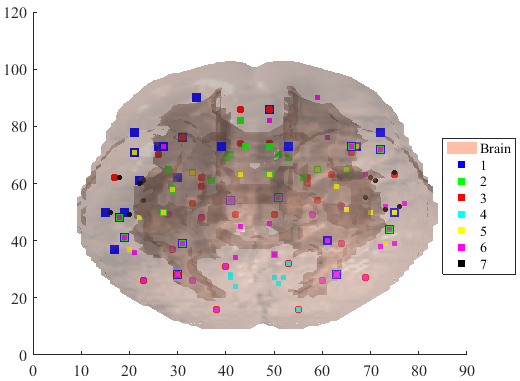}
	}\\
	\subfloat[\ac{MCSE} with $\mathbf{C}^{\text{rest}}$]{
	    \includegraphics[width=2.4in]{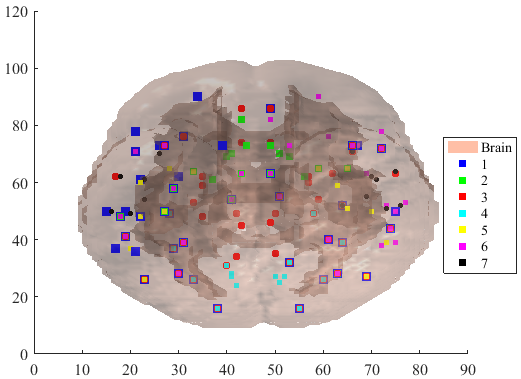}
	}
	\subfloat[kclique]{
	    \includegraphics[width=2.4in]{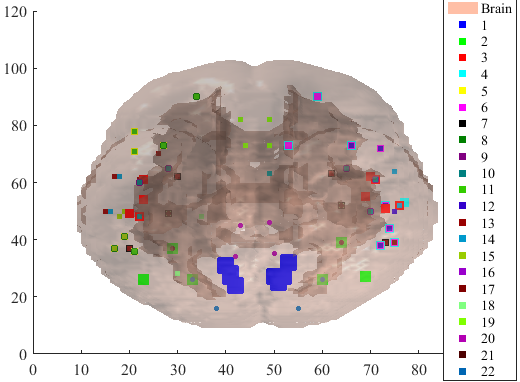}
	}\\
	\subfloat[\ac{SORD}]{
	    \includegraphics[width=2.4in]{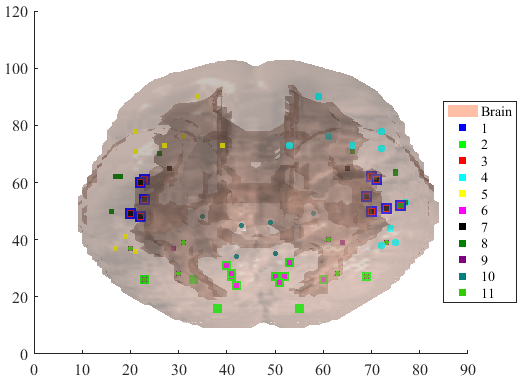}
	}
	\subfloat[\ac{CSORD}]{
	    \includegraphics[width=2.4in]{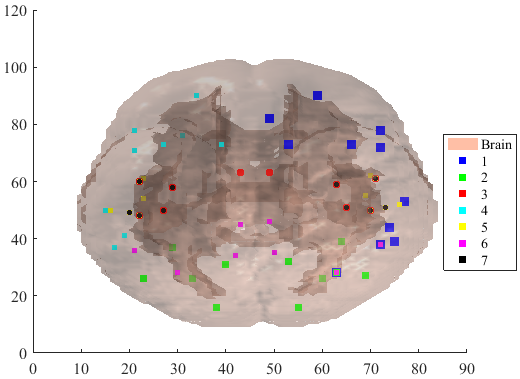}
	}
    \caption{Visualization of Task activation and overlapping subnetworks extracted from our proposed approach and contrasted three other methods. The brain is visualized in the axial view. Our proposed \ac{MCSE} approach outperforms existing state-of-the-art overlapping community detection methods by detecting well-known hubs which reside within subnetwork overlaps.}
    \label{fig:over_bio_methods}
\end{figure}
We further examined the biological meaning of the overlapping subnetworks found using all five methods, \ie our proposed \ac{MCSE} with $\mathbf{C}^{\text{t-r}}$, \ac{MCSE} with $\mathbf{C}^{\text{rest}}$, kclique, \ac{SORD} and \ac{CSORD}, \autoref{fig:over_bio_methods}. We first measured the overlapping ratio by dividing the number of nodes residing in the subnetwork overlaps by the total number of brain regions detected in subnetworks. The ratio of five methods are 0.3482, 0.3482, 0.4444, 0.4328 and 0.2885. Our proposed approach can generate the similar ratio of interacting nodes which reside within subnetwork overlaps to the existing overlapping methods. We note that \ac{CSORD} generated relatively smaller number of interacting nodes, the possible reason is that the strict stability selection resulted in exclusion of some meaningful nodes, which were taken as false detected nodes arising from noise \cite{Yoldemir-2015-Coupled}.

By examining the locations of those interacting nodes, we found that our proposed \ac{MCSE} with $\mathbf{C}^{\text{t-r}}$ approach identified subnetwork overlaps within pre- and postcentral gyri, medial superior frontal cortex, inferior frontal gyrus, superior parietal lobule, precuneous, lateral occipital cortex, occipital pole and frontal orbital cortex; which match well with functional hubs previously identified by graph-theoretical analysis based on the degree of the voxels \cite{Buckner-2009-Cortical}. Besides, brain regions of insula, putamen, thalamus, supramarginal gyrus have been found within subnetwork overlaps, which match well with the connector hubs identified using the centrality measures \cite{Geetharamani-2014-Human}. The results of using \ac{MCSE} with $\mathbf{C}^{\text{rest}}$ is very similar to $\mathbf{C}^{\text{t-r}}$, only that precuneous cortex was missed, and the temporal pole was misclassified into the subnetwork overlaps. This result confirms the benefit of integrating the information from both task and rest data. Both \ac{MCSE} methods also identified 
lingual gyrus and fusiform cortex around as interacting nodes. Lingual gyrus was identified as a hub based on cortical thickness correlation \cite{He-2008-Structural} and the fusiform cortex within occipitotemporal cortex has been found to be intermediary “hub” linking visual and higher linguistic representations \cite{Mano-2012-Role}. 




As for the traditional kclique approach, biologically meaningful subnetwork overlaps were found within inferior frontal gyrus, superior and middle temporal gyri \cite{Buckner-2009-Cortical}, supramarginal gyrus, insula \cite{Geetharamani-2014-Human}, inferior temporal gyrus \cite{Yun-2017-Left}, and occipitotemporal cortex \cite{Mano-2012-Role}. kclique failed to identify all the other aforementioned (connector) hubs which were found using our proposed methods. Instead, regions normally were not considered to reside in subnetwork overlaps were found, such as temporal fusiform cortex, central opercular cortex, and parietal operculum cortex. On the other hand, this kclique approach detected angular gyrus (functioning as a semantic hub) within subnetwork overlaps.

\ac{SORD} was able to find subnetwork overlaps within inferior, superior and middle temporal gyri, superior parietal lobule, lateral occipital cortex, occipital pole and lingual gyrus that match well with functional hubs, but failed to find other hub regions identified by \ac{MCSE}. Instead, \ac{SORD} detected many regions as interacting nodes, which normally are not considered as hubs, such as intracalcarine cortex and cuneal cortex in the visual system, and regions in language related system, including central opercular cortex, parietal operculum cortex, planum polare, planum temporale, heschls gyrus, and supracalcarine cortex.

With relatively lower number of overlapping ratio, \ac{CSORD} identified biological meaningful subnetwork overlaps within regions such as pre- and postcentral gyri, middle temporal gyrus, angular gyrus and lateral occipital cortex, while failed to find any other hubs. Similar to \ac{SORD}, it included some regions in language related system to the subnetwork overlaps, such as central opercular cortex, parietal operculum cortex, and planum temporale. We did not discover the single subnetwork constituting the visual corticostriatal loop, striatothalamo-cortical loop, and cerebello-thalamo-cortical loop, which was found in \cite{Yoldemir-2015-Coupled}. The reason could be this connection was reflected in \ac{AC}, instead of task functional connectivity. 

Collectively, our proposed \ac{MCSE} approach is able to identify subnetwork overlaps which constitute more biologically meaningful brain regions, such as hubs, compared against contrasted methods.
\subsection{Comparison Between the Subnetwork Overlaps and the \ac{RW} Posterior Probability}
\label{sec:multiRW_over}
We also examined the overlapping subnetworks derived from our approach \ac{MCSE} with $\mathbf{C}^{\text{t-r}}$ by assessing the probability of a node belonging to subnetworks using our own recently proposed multimodal \ac{RW} approach \cite{Wang-2017-RW}, to verify that our overlapping subnetwork assignments correspond with the posterior probability. The underlying rationale is that for an interacting node, which resides within the subnetwork overlaps, its probability of belonging to a subnetwork will be distributed across the subnetworks it resides in. On the other hand, an individual node, which does not reside within subnetwork overlaps, would have higher chances to possess a dominant probability of being assigned to a particular subnetwork. Hence the difference of probabilities of a node being assigned to the first two subnetworks with the first two highest probabilities indicates the possibility of a node residing within subnetwork overlaps. Interacting nodes tend to have a smaller value of difference of first two highest probabilities.

We here define the degree of overlapping confidence as the subtraction from one of the difference between the first two highest probabilities of a node being assigned to subnetworks. The nodes identified within the subnetwork overlaps (interacting nodes) and outside of the overlaps (individual nodes) are considered as two populations. For each population, the average overlapping confidence is defined as below in \autoref{eq:prob_conf_interval}:
\begin{equation}
\label{eq:prob_conf_interval}
overConf = \frac{1}{|S|}\sum_{i \in S} (1-(p_i^{max}-p_i^{smax})),
\end{equation}
where $S$ is a set of nodes, either nodes residing within or outside the subnetwork overlaps, and $p^{max}$ is the maximal probability of a node belonging to subnetworks, and $p^{smax}$ is the second maximal probability. Thus, the interacting node population is expected to have higher $overConf$ compared to individual nodes.

We first derived the probabilities of each node being assigned into all possible subnetworks using our recently proposed multimodal \ac{RW} approach \cite{Wang-2017-RW}, where two sources of connectivity matrices are $\mathbf{C}^{\text{rest}}$ and $\bar{\mathbf{C}}^{\text{task}}$, matching with how $\mathbf{C}^{\text{t-r}}$ was generated in our approach. The number of seeds within each subnetwork $n_k$ was set to $[2,\ldots,9]$, where 9 is 75$\%$ of 12, the minimal number of nodes which are included in non-overlapping subnetwork extraction using $\mathbf{C}^{\text{t-r}}$.

\begin{figure}
	\centering
    \includegraphics[width=4.5in]{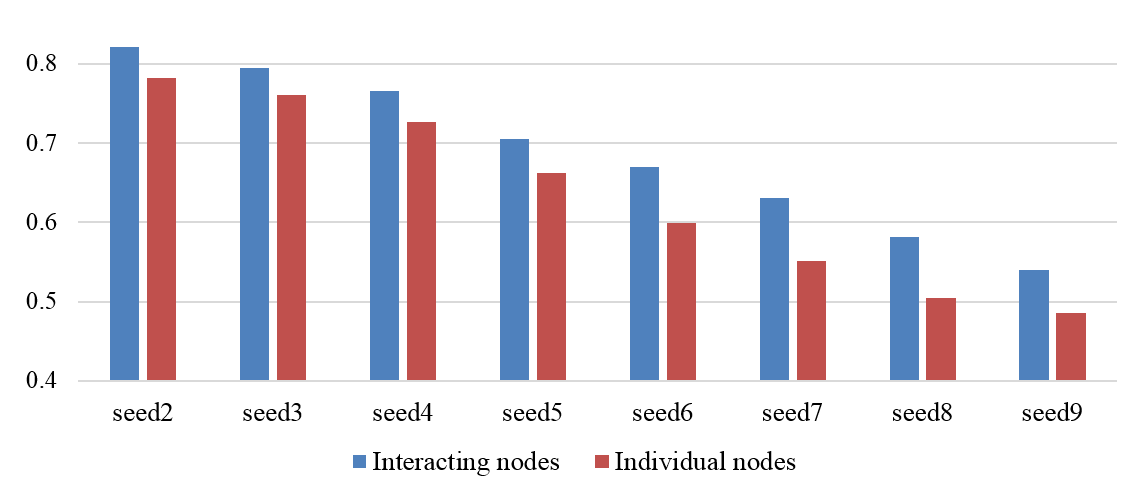}
    \caption{Overlapping confidence of interacting nodes in blue versus indivual nodes in red derived by \ac{MCSE} with $\mathbf{C}^{\text{t-r}}$. The probability of a node being assigned into subnetworks was derived by the \ac{RW} based approach \cite{Wang-2017-RW}.}
    \label{fig:over_level}
\end{figure}

We found that the overlapping confidence of the interacting nodes with an average $overConf$ of 0.6884 are statistically higher than the individual nodes with an average $overConf$ of 0.6338 based on the Wilcoxon signed rank test at p=0.006,
see \autoref{fig:over_level}. This finding confirms that the overlapping subnetwork assignments based on our proposed \ac{MCSE} match with the probability derived independently from our \ac{RW} based approach.  
%
%
\section{Discussion}
\subsection{Benefits of Clique Identification Based on Task Co-activation}
The traditional definition of clique is the fully connected subgraphs identified by the connections between brain regions mostly on resting state connectivity. In our approach, we present a novel way to identify cliques based on the similarity of activation patterns between nodes. We argue that the \textit{clique} concept closely resemble the \textit{canonical network components} that are recruited selectively and repeatedly in different task-induced activities \cite{Park-2014-Graph}. Different from the traditional kclique method \cite{Palla-2005-Uncovering}, our clique-based approach is able to utilize both the task activation information and the task-induced connectivity strength rather than only the binarized connectivity information used in kclique method. Besides, the cliques derived using our approach have flexible clique size, which was determined automatically, having more neuro-scientific justifications than the fixed clique size in \cite{Palla-2005-Uncovering}. Moreover, we estimate the properties from cliques to indicate the importance of cliques, which gives us a better control over falsely including some fake cliques due to noise. We did find cliques within brain areas that well match with hubs, which indicates that our approach can identify subnetwork overlaps with more biological meaning than the traditional kclique method.
\subsection{Multisource Information Integration Improves the Overlapping Subnetwork Extraction}
Compared to the widely used overlapping community detection methods, our approach integrates information from multiple sources. We used both task information (including task activation and connectivity strength) for clique identification and resting state connectivity information for subnetwork membership sharing. The results from reproducibility and biological meaning indicate that our multisource approach, especially \ac{MCSE} with $\mathbf{C}^{\text{t-r}}$, outperforms uni-source methods such as kclique and \ac{SORD}, which has been proven to give better overlapping brain subnetwork extraction results compared to state of the art techniques such as \ac{OSLOM}. We note that our multisource approach further outperformed the multisource version of \ac{SORD}, \ac{CSORD}. The reason could be that clique based idea and the sharing of the node subnetwork membership is more straightforward and have more direct biological intuition than relying on survival probabilities of different genders in evolution, which is used in \ac{CSORD}.
\subsection{Overlapping Subnetwork Identification Corresponds with the \ac{RW} Posterior Probability}
We have identified subnetwork overlaps within brain regions that well match with hubs defined using functional, structural and anatomical information. The results enable us to study the interaction and integration between subnetworks and how interacting nodes (or important hubs) play their roles in the information flow across different subnetworks. We further demonstrated that the assignments of interacting/individual nodes using our proposed \ac{MCSE} correspond with the posterior probability derived independently from our previously proposed \ac{RW} based approach \cite{Wang-2017-RW}. 
The finding of more distinguishable overlapping confidence between two populations of nodes when the number of seeds was set within a range of [6, 8] confirms the merit of using multiple seeds within a reasonable range (not including connector hubs) in the \ac{RW} based approach.
\subsection{Other Considerations}
We have also discovered that the uni-source traditional kclique approach has high computational complexity when the graph density increases, where there exist large number of fully connected subgraphs. The computational complexity of both \ac{SORD} and \ac{CSORD} increases when the bootstrap sampling increases \cite{Yoldemir-2016-Multimodal}. However, the computation time of our proposed \ac{MCSE} is quite reasonable and not sensitive to the graph densities.

In terms of the coverage of the brain area from the subnetwork extraction results, \ac{SORD} and \ac{CSORD} neglected some brain regions which are not selected as significant nodes by the stability selection. While these two approaches offered this extra feature, they sometimes falsely missed important nodes and failed to cover the whole brain for analysis. 
\subsection{Limitations and Future Work}
In this work, we presented an approach to identify cliques based on task information and extract overlapping subnetworks using both task and rest data. However, the ideal multimodal framework would be able to integrate \ac{AC} into the fusion for detecting overlapping subnetworks. The challenge is to discover the relationship between \ac{AC} and task activation, which enables the clique identification to incorporate anatomical information. Our future work will focus on integrating \ac{AC} into task-activation based clique identification, or into the multimodal subnetwork membership assignment, \eg using the multimodal \ac{RW} approach or the multislice approach \cite{Mucha-2010-Community}.
\section{Conclusion}
We proposed an approach for multisource overlapping brain subnetwork extraction using canonical network components, \ie cliques, which we defined based on task co-activation. Based on the clique concept, we investigated overlapping subnetworks based on a label sharing scheme which incorporates the rest data information and task data embedded with higher order relations.
We have demonstrated that integrating multimodal/multisource information and using high order relations result in better subnetwork extraction in terms of the overlaps to well-established brain systems, test-retest repeatability, inter-subject reproducibility and biological meaning.
\section{List of Acronyms}

\begin{acronym}
\acro{AAL}{Automated Anatomical Labeling}
\acro{AC}{Anatomical Connectivity}
\acro{ACC}{Anterior Cingulate Cortex}
\acro{AD}{Alzheimer’s disease}
\acro{BOLD}{Blood Oxygenated Level Dependent}
\acro{BP}{Boundary Point}
\acro{CBP}{connectivity based parcellation}
\acro{Cg}{Cingulate Cortex}
\acro{CIS}{Connected Iterative Scan}
\acro{CNN}{Convolutional Neural Networks}
\acro{COREG}{CO-training with REGularization}
\acro{CPM}{Clique Percolation Method}
\acro{CSA}{Constant Solid Angle}
\acro{CSORD}{Coupled Stable Overlapping Replicator Dynamics}
\acro{DAE}{Deep Auto-Encoder}
\acro{DALYs}{Disability-Adjusted Life Years}
\acro{DBN}{Deep Belief Network} 
\acro{DC}{physical distance}
\acro{DMN}{Default Mode Network}
\acro{dMRI}{Diffusion-weighted Magnetic Resonance Imaging}
\acro{DNNs}{Deep Neural Networks}
\acro{DOF}{degree of freedom}
\acro{DSC}{Dice Similarity Coefficient}
\acro{DTI}{Diffusion Tensor Imaging}
\acro{EC}{Effective Connectivity}
\acro{ECN}{Executive Control Network}
\acro{EEG}{Electroencephalography} 
\acro{EPI}{Echo-Planar Imaging}
\acro{FA}{Fractional Anisotropy}
\acro{FC}{Functional Connectivity}
\acro{fMRI}{Functional Magnetic Resonance Imaging}
\acro{GBD}{Global Burden of Disease}
\acro{GLM}{General Linear Model}
\acro{GMM}{Gaussian Mixture Model}
\acro{GROUSE}{Grassmannian Rank-One Update Subspace Estimation} 
\acro{GT}{Global Thresholding}
\acro{HARDI}{High Angular Resolution Diffusion Imaging}
\acro{HCP}{Human Connectome Project}
\acro{HO}{Harvard-Oxford}
\acro{ICA}{Independent Component Analysis}
\acro{ICC}{Intra-Class Correlation}
\acro{ICNs}{Intrinsic Connectivity Networks}
\acro{IP}{Interior Point}
\acro{IPL}{Inferior Parietal Lobule}
\acro{IQ}{Intelligence Quotient}
\acro{LECN}{Left Executive Control Network}
\acro{LMaFit}{Low-Rank Matrix Fitting}
\acro{LT}{Local Thresholding}
\acro{MC}{Multimodal Connectivity} 
\acro{MCNF}{Matrix Completion with Nonnegative Factorization}
\acro{MCSE}{Multisource Clique-based Subnetwork Extraction}
\acro{MEG}{Magnetoencephalography}
\acro{MITK}{Medical Imaging Interaction Toolkit}
\acro{mmRW}{multi-modal Random Walker}
\acro{MNI}{Montreal Neurological Institute}
\acro{MRI}{Magnetic Resonance Imaging}
\acro{MST-KNN}{minimal spanning tree and k-nearest neighbors}
\acro{MVSC}{MultiView Spectral Clustering}
\acro{Ncuts}{Normalized cuts}
\acro{NMD}{Neighborhood-information-embedded Multiple Density}
\acro{NMF}{Non-negative Matrix Factorization}
\acro{NMI}{Normalized Mutual Information}
\acro{NP}{Non-deterministic Polynomial-time}
\acro{NRMSE}{normalized root-mean-squared-error}
\acro{ODF}{Orientation Distribution Function}
\acro{OSLOM}{Order Statistics Local Optimization Method} 
\acro{PCA}{Principal Component Analysis}
\acro{PCC}{Posterior Cingulate Cortex}
\acro{PD}{Parkinson’s disease}
\acro{PDD}{Principal Diffusion Direction}
\acro{PDF}{Probability Density Function}
\acro{PET}{Positron Emission Tomography}
\acro{RBM}{Restricted Boltzmann Machines}
\acro{RD}{Replicator Dynamics}
\acro{RM}{random parcellation}
\acro{ROI}{region of interest}
\acro{RQ}{Research Questions}
\acro{rs-fcMRI}{Resting State Functional Connectivity based on MRI}
\acro{RW}{Random Walker}
\acro{SAE}{Stacked Auto-Encoder}
\acro{sMRI}{Structural Magnetic Resonance Imaging}
\acro{SNR}{signal-to-noise ratio}
\acro{SORD}{Stable Overlapping Replicator Dynamics}
\acro{SVR}{Support Vector Regression}
\acro{t-fcMRI}{Task Functional Connectivity based on MRI}
\acro{TR}{repetition time}
\acro{WHO}{World Health Organization}
\acro{3D}{three-dimensional}
\end{acronym}

%

\bibliographystyle{splncs}
\bibliography{biblio}
\end{document}